\documentclass[preprintnumbers, prd, showpacs, floatfix, onecolumn,
preprintnumbers, letterpaper, superscriptaddress,nofootinbib]{revtex4}
\usepackage{amsfonts}
\usepackage{amsmath}
\usepackage{amssymb,epsf}
\usepackage{latexsym}
\usepackage{graphicx,epsfig}
\usepackage{amssymb}
\usepackage{subfigure}
\usepackage{epstopdf}
\usepackage[colorlinks=true,citecolor=blue,linkcolor=blue,urlcolor=black]{hyperref}

\graphicspath{{Images/}}
\begin{document}

\title{Holographic conductivity in the massive gravity with power-law
Maxwell field}
\author{A. Dehyadegari}
\email{adehyadegari@shirazu.ac.ir}
\affiliation{Physics Department and Biruni Observatory, Shiraz University, Shiraz 71454,
Iran}
\author{M. Kord Zangeneh}
\email{kordzangeneh.mehdi@gmail.com}
\affiliation{Physics Department, Faculty of Science, Shahid Chamran University of Ahvaz,
Ahvaz 61357-43135, Iran}
\affiliation{Center of Astronomy and Astrophysics, Department of Physics and Astronomy,
Shanghai Jiao Tong University, Shanghai 200240, China}
\author{A. Sheykhi}
\email{asheykhi@shirazu.ac.ir}
\affiliation{Physics Department and Biruni Observatory, Shiraz University, Shiraz 71454,
Iran}
\affiliation{Research Institute for Astrophysics and Astronomy of Maragha (RIAAM), P.O.
Box 55134-441, Maragha, Iran}

\begin{abstract}
We obtain a new class of topological black hole solutions in $(n+1)$%
-dimensional massive gravity in the presence of the power-Maxwell
electrodynamics. We calculate the conserved and thermodynamic quantities of
the system and show that the first law of thermodynamics is satisfied on the
horizon. Then, we investigate the holographic conductivity for the four and
five dimensional black brane solutions. For completeness, we study the
holographic conductivity for both massless ($m=0$) and massive ($m\neq 0$)
gravities with power-Maxwell field. The massless gravity enjoys
translational symmetry whereas the massive gravity violates it. For massless
gravity, we observe that the real part of conductivity, $\mathrm{Re}[\sigma
] $, decreases as charge $q$ increases when frequency $\omega $ tends to
zero, while the imaginary part of conductivity, $\mathrm{Im}[\sigma ]$,
diverges as $\omega \rightarrow 0$. For the massive gravity, we find that $%
\mathrm{Im}[\sigma ]$ is zero at $\omega =0$ and becomes larger as $q$\
increases (temperature decreases), which is in contrast to the massless
gravity. It also has a maximum value for $\omega \neq 0$ which increases
with increasing $q$ (with fixed $p$) or increasing $p$ (with fixed $q$) for (%
$2+1$)-dimensional dual system, where $p$ is the power parameter of the
power-law Maxwell field. Interestingly, we observe that in contrast to the
massless case, $\mathrm{Re}[\sigma ]$ has a maximum value at $\omega =0$
(known as the Drude peak) for $p=\left( n+1\right) /4$ (conformally
invariant electrodynamics) and this maximum increases with increasing $q$.
In this case ($m\neq 0$) and for different values of $p$, the real and
imaginary parts of the conductivity has a relative extremum for $\omega \neq
0$. Finally, we show that for high frequencies, the real part of the
holographic conductivity have the power law behavior in terms of frequency, $%
\omega ^{a}$ where $a\propto (n+1-4p)$. Some similar behaviors for high
frequencies in possible dual CFT systems have been reported in experimental
observations.
\end{abstract}

\pacs{97.60.Lf, 04.70.-s, 71.10.-w, 04.70.Bw, 04.30.-w. }
\maketitle

\section{Introduction}

A century after Einstein's discovery namely general relativity, the domain
of its applications has become as vast as it covers even condensed matter
physics which seemed at the opposite end of physics building compared to
gravity \cite{HDCMP}. This strange topic which connects gravity to almost
all fields of physics (see \cite{Nat}) is called gauge/gravity duality
(GGD); the extended version of AdS/CFT correspondence \cite{MW}. GGD has
attracted increasing interests during recent years and become one of the
most promising fields of physics which is hoped to be able to solve many of
unsolved problems in different fields of physics including condensed matter
physics.

Real materials in condensed matter physics do not respect the translational
symmetry i.e. there is a dissipation in momentum. The momentum dissipation
may come from the existence of a lattice or impurities. Although this
dissipation has no important influence on the values of some observable, it
affects the behavior of some others for instance conductivity. The DC
conductivity in the presence of translational symmetry diverges, whereas in
the absence of this symmetry (when momentum is dissipating) it has a finite
value. In the context of GGD, it is important to study a gravity model which
includes holographic momentum dissipation. There are some attempts to
construct such gravity model \cite{momdis}. One of these models proposed by
D. Vegh \cite{vegh}, provides an effective bulk description of a theory in
which momentum is no longer conserved. The conservation of momentum is due
to the diffeomorphism invariance of stress-energy tensor in dual theory. In 
\cite{vegh}, the proposal is to break this symmetry holographically by
giving a mass to graviton state. The resulting gravity is therefore \textit{%
massive gravity}. One of the advantages of this theory is that the black
hole solutions of it are solvable analytically and therefore it is an
excellent toy model to study holographically the properties of materials
without momentum conservation.

Thermal behaviors of black hole solutions in the context of massive gravity
was explored extensively in recent years \cite{vegh,cai,hendimann,thermo}.
Thermodynamics of linearly charged massive black branes has been
investigated in \cite{vegh}. In \cite{cai}, a class of higher-dimensional
linearly charged solutions with positive, negative and zero constant
curvature of horizon in the context of massive gravity accompanied by a
negative cosmological constant has been presented and thermodynamics and
phase structure of these black solutions have been studied in both canonical
and grand canonical ensembles. In \cite{hendimann}, van der Waals phase
transitions of linearly charged black holes in massive gravity have been
investigated and it has been shown that the massive gravity can present
substantially different thermodynamic behavior in comparison with Einstein
gravity. Also it has been shown that the graviton mass can cause a range of
new phase transitions for topological black holes which are forbidden for
other cases. The properties of massive solutions have been studied in
different scenarios \cite{dsMG}. From holographic point of view, the
behaviors of different holographic quantities have been studied \cite%
{vegh,holo,1404.5321,1407.0306,1512.07035,1611.00677,1504.00535,1507.03105,1704.03989,matteo1,matteo2,matteo3,matteo4,1612.03627}%
. The behavior of holographic conductivity for systems dual to linearly
charged massive black branes has been explored in \cite{vegh}. In \cite%
{1404.5321}, a holographic superconductor has been constructed in the
massive gravity background. \cite{1512.07035} studies holographic
superconductor-normal metal-superconductor Josephon junction in the massive
gravity. Also the holographic thermalization process has been investigated
in this context \cite{1611.00677}. Analytic DC thermo-electric
conductivities in the context of massive gravity have been calculated in 
\cite{1407.0306}. In massive Einstein-Maxwell-dilaton gravity, DC and Hall
conductivities have been computed in \cite{1504.00535}. \cite{1507.03105}
presents a holographic model for insulator/metal phase transition and
colossal magnetoresistance within massive gravity. Inspired by the recent
action/complexity duality conjecture, it has been shown in \cite{1612.03627}
that the holographic complexity grows linearly with time in the context of
massive gravity.

As we mentioned above, one of the quantities which is affected by momentum
dissipation is conductivity. On the other hand, the choice of
electrodynamics model has a direct influence on the behavior of
conductivity. So, it is worthy to consider the effects of nonlinearity as
well as massive gravity on the conductivity of the black hole solutions. It
is well-known that the nonlinear electrodynamics brings reach physics
compared to the linear Maxwell electrodynamics. For example, Maxwell theory
is conformally invariant only in four dimensions and thus the corresponding
energy-momentum tensor is only traceless in four dimensions. A natural
question then arises: Is there an extension of Maxwell action in arbitrary
dimensions that is traceless and hence possesses the conformal invariance?
The answer is positive and the invariant Maxwell action under conformal
transformation $g_{\mu \nu }\rightarrow \Omega ^{2}g_{\mu \nu }$, $A_{\mu
}\rightarrow A_{\mu }$ in $(n+1)$-dimensions is given by \cite{PLM}, 
\begin{equation*}
S_{m}=\int {d^{n+1}x\sqrt{-g}(-\mathcal{F})^{p}},
\end{equation*}%
where $\mathcal{F}=F_{\mu \nu}F^{\mu \nu}$ is the Maxwell
invariant, provided $p=(n+1)/4$. The associated energy-momentum tensor of
the above Maxwell action is given by%
\begin{equation}
T_{\mu \nu }=2\left( pF_{\mu \eta }F_{\nu }^{\text{ }\eta }\mathcal{F}^{p-1}-%
\frac{1}{4}g_{\mu \nu }\mathcal{F}^{p}\right) .  \label{T}
\end{equation}%
One can easily check that the above energy-momentum tensor is traceless for $%
p=(n+1)/4$. Also, quantum electrodynamics predicts that the electrodynamic
field behaves nonlinearly through the presence of virtual charged particles
that is reported by Heisenberg and Euler \cite{HE}.\ Hence, nonlinear
electrodynamics has been subject of much researches \cite{NEM1,NEM2,NEM3}.
This motivates us to extend the linearly charged black hole solutions of
massive gravity \cite{vegh,cai} to nonlinearly charged ones in the presence
of power-law Maxwell electrodynamics and investigate the thermodynamics of
them as well as the behavior of conductivity corresponding to the dual
system.\ In addition to power-law Maxwell electrodynamics, other types of
nonlinear electrodynamics have been introduced in \cite{BI,Selong,hendibtz}.
In spite of the special property for $p=(n+1)/4$, different aspects of
various solutions have been investigated for different $p$'s \cite%
{NFP,hendi,shey1}. In the context of AdS/CFT correspondence, the power-law Maxwell field
has been considered as electrodynamics source in \cite%
{holsup1,holsup2,holsup3,shey200,shey201,shey202}.

The layout of this letter is as follows. In section \ref{lifsol}, we present
the action of the massive gravity in the presence of power-Maxwell
electrodynamics and then by varying the action we obtain the field
equations. We also derive a class of topological black hole solutions of the
field equations in higher dimensions. In section \ref{Therm}, we study
thermodynamics of the solutions and examine the first law of thermodynamics
for massive black holes with power-law Maxwell field. In section \ref%
{conductivity}, we investigate the holographic conductivity of black brane
solutions in the presence of a power-law Maxwell gauge field. In particular,
we shall disclose the effects of the power-law Maxwell electrodynamics as
well as massive gravity on the holographic conductivity of dual systems. We
finish with closing remarks in section \ref{Clos}.

\section{Action and massive gravity solutions\label{lifsol}}

\begin{figure*}[t]
\begin{center}
\begin{minipage}[b]{0.32\textwidth}\begin{center}
       \subfigure[~$k=0$, $q_{\rm ext}=2.03$]{
                \label{fig1a}\includegraphics[width=\textwidth]{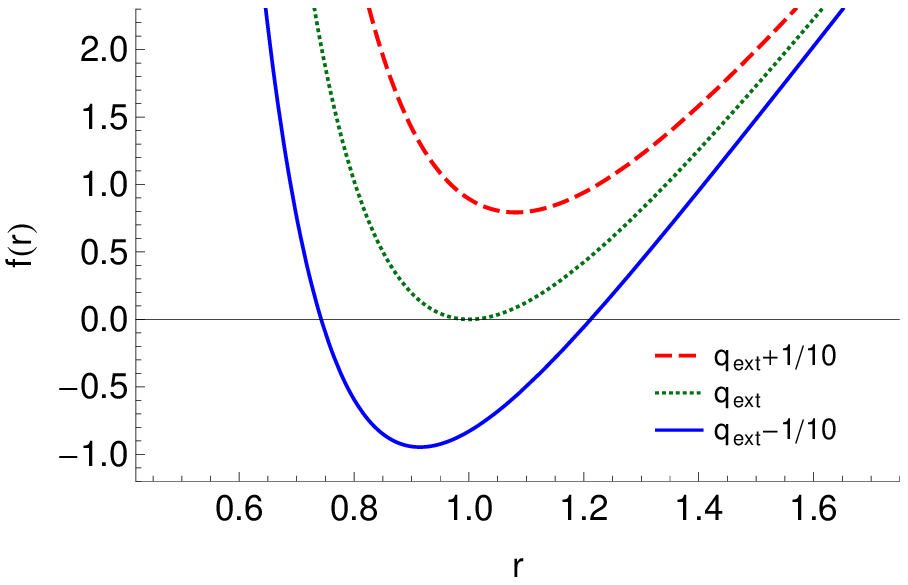}\qquad}
    \end{center}\end{minipage} \hskip+0cm 
\begin{minipage}[b]{0.32\textwidth}\begin{center}
        \subfigure[~$k=1$, $q_{\rm ext}=2.25$]{
                 \label{fig1b}\includegraphics[width=\textwidth]{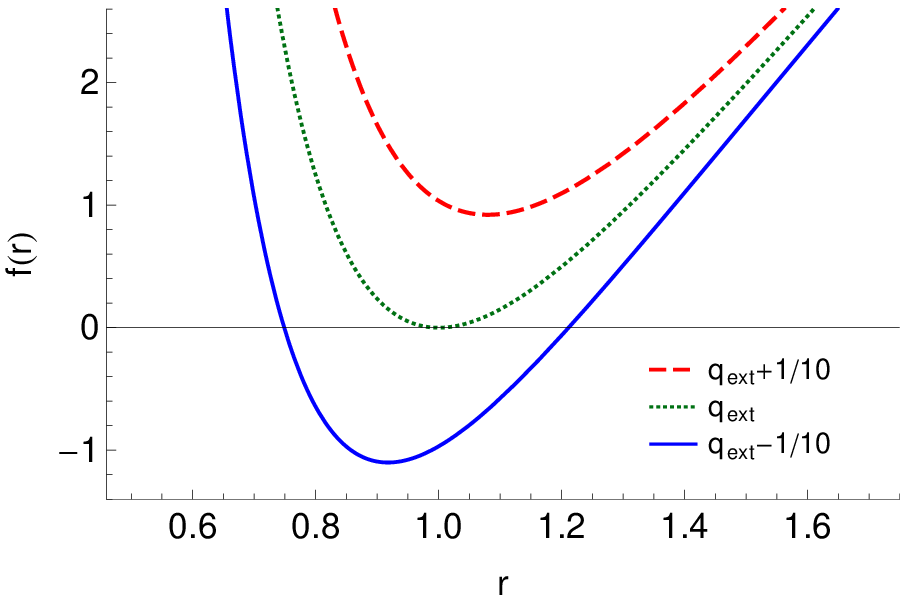}\qquad}
    \end{center}\end{minipage} \hskip0cm 
\begin{minipage}[b]{0.32\textwidth}\begin{center}
         \subfigure[~$k=-1$, $q_{\rm ext}=1.78$]{
                  \label{fig1c}\includegraphics[width=\textwidth]{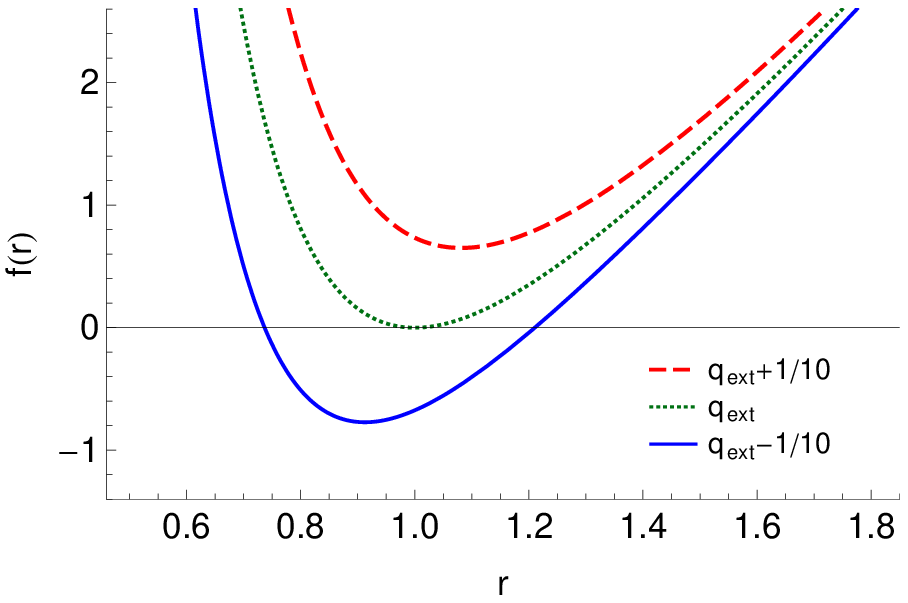}\qquad}
    \end{center}\end{minipage} \hskip0cm
\end{center}
\caption{The behavior of $f(r)$ versus $r$ for $n=4$, $l=1$, $p=5/4$, $m=1$, 
$r_{+}=1$, $c_{0}=1$, $c_{1}=1$, $c_{2}=3/2$, $c_{3}=-1/2$ and $c_{4}=1$.}
\label{fig1}
\end{figure*}

The ($n+1$)-dimensional ($n\geq 3$) action describing Einstein-massive
gravity accompanied by a negative cosmological constant $\Lambda $ in the
presence of power-law Maxwell electrodynamics is%
\begin{eqnarray}
\mathcal{S} &=&\int d^{n+1}x\mathcal{L},  \label{Action} \\
\mathcal{L} &=&\frac{\sqrt{-g}}{16\pi }\left[ \mathcal{R}-2\Lambda +\left( -%
\mathcal{F}\right) ^{p}+m^{2}\sum_{i}^{4}c_{i}\mathcal{U}_{i}(g,\Gamma )%
\right] ,  \label{DL}
\end{eqnarray}%
where $g$ and $\mathcal{R}$ are respectively the determinant of the metric
and the Ricci scalar and $\Lambda =-n(n-1)/2l^{2}$ is the negative
cosmological constant where $l$ is the AdS radius. $\mathcal{F}=F_{\mu \nu
}F^{\mu \nu }$ and $F_{\mu \nu }=\partial _{\lbrack \mu }A_{\nu ]}$ is
electrodynamic tensor where $A_{\nu }$ is vector potential. $p$ determines
the nonlinearity of the electrodynamic field. For $p=1$, the linear Maxwell
gauge field will be recovered. In action (\ref{Action}), $\Gamma $ is the
reference metric, $c_{i}$'s are constants and $\mathcal{U}_{i}$'s are
symmetric polynomials of eigenvalues of the $(n+1)\times (n+1)$ matrix $%
\mathcal{K}_{\nu }^{\mu }\equiv \sqrt{g^{\mu \alpha }\Gamma _{\alpha \nu }}$
so that%
\begin{eqnarray}
\mathcal{U}_{1} &=&\left[ \mathcal{K}\right] , \\
\mathcal{U}_{2} &=&\left[ \mathcal{K}\right] ^{2}-\left[ \mathcal{K}^{2}%
\right] , \\
\mathcal{U}_{3} &=&\left[ \mathcal{K}\right] ^{3}-3\left[ \mathcal{K}\right] %
\left[ \mathcal{K}^{2}\right] +2\left[ \mathcal{K}^{3}\right] , \\
\mathcal{U}_{4} &=&\left[ \mathcal{K}\right] ^{4}-6\left[ \mathcal{K}^{2}%
\right] \left[ \mathcal{K}\right] ^{2}+8\left[ \mathcal{K}^{3}\right] \left[ 
\mathcal{K}\right] +3\left[ \mathcal{K}^{2}\right] ^{2}-6\left[ \mathcal{K}%
^{4}\right] ,
\end{eqnarray}%
where the square root in $\mathcal{K}$ is related to mean matrix square root
i.e. $\left( \sqrt{\mathcal{K}}\right) _{\nu }^{\mu }\left( \sqrt{\mathcal{K}%
}\right) _{\lambda }^{\nu }=\mathcal{K}_{\lambda }^{\mu }$ and rectangular
brackets mean trace $\left[ \mathcal{K}\right] \equiv \mathcal{K}_{\mu
}^{\mu }$. Here $m$ is the massive gravity parameter so that in limit $%
m\rightarrow 0$, one recovers the diffeomorphism invariant Einstein-Hilbert
action with a gauge field and a negative cosmological constant. The
equations of motion for gravitation and gauge field are%
\begin{equation}
R_{\mu \nu }-\frac{1}{2}\mathcal{R}g_{\mu \nu }+\Lambda g_{\mu \nu
}-2pF_{\mu \lambda }F_{\nu }^{\text{ \ }\lambda }\left( -\mathcal{F}\right)
^{p-1}-\frac{1}{2}\left( -\mathcal{F}\right) ^{p}g_{\mu \nu }+m^{2}\chi
_{\mu \nu }=0,  \label{Field equation}
\end{equation}%
\begin{equation}
\nabla _{\mu }\left( \mathcal{F}^{p-1}F^{\mu \nu }\right) =0,
\label{Maxwell equation}
\end{equation}%
which are obtained by varying the action (\ref{Action}) with respect to the
metric tensor $g_{\mu \nu }$ and gauge field $A_{\mu }$ respectively. In Eq.
(\ref{Field equation}), we have%
\begin{eqnarray}
\chi _{\mu \nu } &=&-\frac{c_{1}}{2}\left( \mathcal{U}_{1}g_{\mu \nu }-%
\mathcal{K}_{\mu \nu }\right) -\frac{c_{2}}{2}\left( \mathcal{U}_{2}g_{\mu
\nu }-2\mathcal{U}_{1}\mathcal{K}_{\mu \nu }+2\mathcal{K}_{\mu \nu
}^{2}\right) -\frac{c_{3}}{2}(\mathcal{U}_{3}g_{\mu \nu }-3\mathcal{U}_{2}%
\mathcal{K}_{\mu \nu }  \notag \\
&&+6\mathcal{U}_{1}\mathcal{K}_{\mu \nu }^{2}-6\mathcal{K}_{\mu \nu }^{3})-%
\frac{c_{4}}{2}(\mathcal{U}_{4}g_{\mu \nu }-4\mathcal{U}_{3}\mathcal{K}_{\mu
\nu }+12\mathcal{U}_{2}\mathcal{K}_{\mu \nu }^{2}-24\mathcal{U}_{1}\mathcal{K%
}_{\mu \nu }^{3}+24\mathcal{K}_{\mu \nu }^{4}).
\end{eqnarray}%
The static spacetime line element takes the usual form%
\begin{equation}
ds^{2}=-f(r)dt^{2}+f^{-1}(r)dr^{2}+r^{2}h_{ij}dx^{i}dx^{j},  \label{Metric}
\end{equation}%
where $f(r)$ is the metric function and $h_{ij}$ is a function of
coordinates $x_{i}$ which spanned an $(n-1)$-dimensional hypersurface with
constant scalar curvature $(n-1)(n-2)k$ and volume $\omega _{n-1}$. Without
loss of generality, one can take $k=0,1,-1$, such that the black hole
horizon or cosmological horizon in (\ref{Metric}) can be a zero (flat),
positive (elliptic) or negative (hyperbolic) constant curvature
hypersurface. The reference metric (fixed symmetric tensor) $\Gamma _{\mu
\nu }$ can be considered as \cite{vegh,cai}%
\begin{equation}
\Gamma _{\mu \nu }=\mathrm{diag}(0,0,c_{0}^{2}h_{ij}),  \label{f11}
\end{equation}%
where $c_{0}$ is a positive constant. Using (\ref{Metric}) and (\ref{f11}),
one can easily calculates $\mathcal{U}_{i}$'s as%
\begin{eqnarray}
\mathcal{U}_{1} &=&\frac{(n-1)c_{0}}{r},  \notag \\
\mathcal{U}_{2} &=&\frac{(n-1)(n-2)c_{0}^{2}}{r^{2}},  \notag \\
\mathcal{U}_{3} &=&\frac{(n-1)(n-2)(n-3)c_{0}^{3}}{r^{3}},  \notag \\
\mathcal{U}_{4} &=&\frac{(n-1)(n-2)(n-3)(n-4)c_{0}^{4}}{r^{4}}.  \label{PL}
\end{eqnarray}%
Notice that $\mathcal{U}_{3}$ and $\mathcal{U}_{4}$ vanish for ($3+1$%
)-dimensional spacetime while $\mathcal{U}_{4}=0$ for ($4+1$)-dimensional
spacetime. Using the metric (\ref{Metric}), the electrodynamic field can be
immediately found as 
\begin{equation}
F_{tr}=-F_{rt}=\frac{q}{r^{\left( n-1\right) /\left( 2p-1\right) }},
\label{Gaugefield}
\end{equation}%
where $q$ is a constant parameter related to the total charge of black hole.
Inserting Eqs. (\ref{f11}), (\ref{PL}) and (\ref{Gaugefield}) into field
equations (\ref{Field equation}), one receives%
\begin{eqnarray}
\frac{f^{\prime }}{r}+\frac{(n-2)f}{r^{2}}-\frac{(n-2)k}{r^{2}}+\frac{%
2\Lambda }{n-1}+\frac{2p-1}{n-1}\left( 2q^{2}r^{-\frac{2n-2}{2p-1}}\right)
^{p}-\frac{c_{0}m^{2}}{r}\left( c_{1}+\frac{(n-2)c_{0}c_{2}}{r}\right. && 
\notag \\
\left. +\frac{(n-2)(n-3)c_{0}^{2}c_{3}}{r^{2}}+\frac{%
(n-2)(n-3)(n-4)c_{0}^{3}c_{4}}{r^{3}}\right) &=&0,  \label{Eq1}
\end{eqnarray}

\begin{eqnarray}
f^{\prime \prime }+\frac{2(n-2)f^{\prime }}{r}+\frac{(n-2)(n-3)f}{r^{2}}-%
\frac{(n-2)(n-3)k}{r^{2}}+2\Lambda -\left( 2q^{2}r^{-\frac{2n-2}{2p-1}%
}\right) ^{p}-\frac{(n-2)c_{0}m^{2}}{r}\left( c_{1}+\frac{(n-3)c_{0}c_{2}}{r}%
\right. &&  \notag \\
\left. +\frac{(n-3)(n-4)c_{0}^{2}c_{3}}{r^{2}}+\frac{%
(n-3)(n-4)(n-5)c_{0}^{3}c_{4}}{r^{3}}\right) &=&0,  \notag \\
&&  \label{Eq2}
\end{eqnarray}%
where prime denotes the derivative with respect to $r$. Solving above
equations, $f(r)$ can be obtained as 
\begin{eqnarray}
f(r) &=&k-\frac{m_{0}}{r^{n-2}}-\frac{2\Lambda r^{2}}{n(n-1)}+\frac{%
2^{p}q^{2p}(2p-1)^{2}}{(n-1)(n-2p)r^{2(np-3p+1)/(2p-1)}}  \notag \\
&&+\frac{c_{0}m^{2}r}{n-1}\left( c_{1}+\frac{(n-1)c_{0}c_{2}}{r}+\frac{%
(n-1)(n-2)c_{0}^{2}c_{3}}{r^{2}}+\frac{(n-1)(n-2)(n-3)c_{0}^{3}c_{4}}{r^{3}}%
\right) ,  \label{Metricfunction}
\end{eqnarray}%
\qquad where $m_{0}$ is an integration constant which is related to total
mass of black hole as we see later. One may note that the metric function (%
\ref{Metricfunction}) reduces to those of Refs. \cite{vegh,cai} in the case $%
p=1$. Also the solution (\ref{Metricfunction}), in the absent of massive
parameter ($m=0$), leads to%
\begin{equation}
f_{0}(r)=k-\frac{m_{0}}{r^{n-2}}-\frac{2\Lambda r^{2}}{n(n-1)}+\frac{%
2^{p}q^{2p}(2p-1)^{2}}{(n-1)(n-2p)r^{2(np-3p+1)/(2p-1)}},  \label{f0}
\end{equation}%
which was presented in \cite{hendi}. The mass parameter ($m_{0}$) in Eq. (%
\ref{Metricfunction}) can be found as%
\begin{eqnarray}
m_{0} &=&kr_{+}^{n-2}-\frac{2\Lambda r_{+}^{n}}{n(n-1)}+\frac{%
2^{p}q^{2p}(2p-1)^{2}}{(n-1)(n-2p)r_{+}^{(n-2p)/(2p-1)}}  \notag \\
&&+\frac{c_{0}m^{2}r_{+}^{n-1}}{n-1}\left( c_{1}+\frac{(n-1)c_{0}c_{2}}{r_{+}%
}+\frac{(n-1)(n-2)c_{0}^{2}c_{3}}{r_{+}^{2}}+\frac{%
(n-1)(n-2)(n-3)c_{0}^{3}c_{4}}{r_{+}^{3}}\right) ,  \label{ZM}
\end{eqnarray}%
where $r_{+}$ is the radius of the event horizon given by the largest root
of $f(r_{+})=0$. According to Eq. (\ref{Gaugefield}) and regarding $%
A_{t}(r)=\int F_{rt}dr$, the gauge potential $A_{t}$ can be calculated as%
\begin{equation}
A_{t}\left( r\right) =\mu +\frac{q(2p-1)}{(n-2p)r^{(n-2p)/(2p-1)}}.
\label{potential}
\end{equation}%
In (\ref{potential}), $\mu $ is the chemical potential of the quantum field
theory locates on boundary which can be found by demanding the regularity
condition on the horizon i.e. $A_{t}\left( r_{+}\right) =0$ as%
\begin{equation}
\mu =\frac{q(2p-1)}{(2p-n)r_{+}^{(n-2p)/(2p-1)}}.
\end{equation}%
One should note that the electric potential $A_{t}\left( r\right) $ has a
finite value at infinity ($r\rightarrow \infty $) provided the parameter $p$
is restricted as%
\begin{equation}
\frac{1}{2}<p<\frac{n}{2},
\end{equation}%
obtained from $(n-2p)/(2p-1)>0$. One can also obtain the electric potential
as%
\begin{equation}
U=A_{\nu }\chi ^{\nu }\left\vert _{r\rightarrow ref}-A_{\nu }\chi ^{\nu
}\right\vert _{r=r_{+}},  \label{Pot}
\end{equation}%
where $\chi =C\partial _{t}$ is the null generator of the horizon and $C$ is
a constant. When one applies the power-law Maxwell electrodynamics, it is
common to use a general Killing vector with a constant $C$\ \cite{14,15}.
This is due to the fact that every linear combination of Killing vectors is
also a Killing vector. Then, $C$ is fixed so that the first law of
thermodynamics is satisfied \cite{14,15}. For linear Maxwell case ($p=1$),
the constant $C$ reduces to $1$. Choosing infinity as the reference point,
one can calculate the electric potential energy%
\begin{equation}
U=C\mu .  \label{Poten}
\end{equation}%
One can obtain the Hawking temperature of the black hole on the event
horizon as%
\begin{eqnarray}
T &=&\frac{f^{\prime }\left( r_{+}\right) }{4\pi }  \notag \\
&=&\frac{(n-2)k}{4\pi r_{+}}-\frac{2\Lambda r_{+}}{4\pi (n-1)}+\frac{%
2^{p}q^{2p}(1-2p)}{4\pi (n-1)r_{+}^{(2p\left[ n-2\right] +1)/(2p-1)}}  \notag
\\
&&+\frac{c_{0}m^{2}}{4\pi }\left( c_{1}+\frac{(n-2)c_{0}c_{2}}{r_{+}}+\frac{%
(n-2)(n-3)c_{0}^{2}c_{3}}{r_{+}^{2}}+\frac{(n-2)(n-3)(n-4)c_{0}^{3}c_{4}}{%
r_{+}^{3}}\right) .
\end{eqnarray}%
The extremal black hole, whose temperature vanishes, can be also determined
by an extremal charge, 
\begin{eqnarray}
q_{\mathrm{ext}}^{2p} &=&\frac{(n-1)(n-2)r_{\mathrm{ext}}^{2\left[ p(n-3)+1%
\right] /(2p-1)}}{(2p-1)2^{p}}-\frac{\Lambda r_{\mathrm{ext}%
}^{2p(n-1)/(2p-1)}}{(2p-1)2^{p-1}}  \notag \\
&&+\frac{c_{0}m^{2}(n-1)r_{\mathrm{ext}}^{\left[ 2p(n-2)+1\right] /(2p-1)}}{%
(2p-1)2^{p}}\left( c_{1}+\frac{(n-2)c_{0}c_{2}}{r_{\mathrm{ext}}}+\frac{%
(n-2)(n-3)c_{0}^{2}c_{3}}{r_{\mathrm{ext}}^{2}}+\frac{%
(n-2)(n-3)(n-4)c_{0}^{3}c_{4}}{r_{\mathrm{ext}}^{3}}\right) ,  \notag \\
&&
\end{eqnarray}%
For $q>q_{\mathrm{ext}}$, there is a naked singularity in spacetime while $%
q<q_{\mathrm{ext}}$ describes solutions with two inner and outer horizons ($%
r_{+}$ and $r_{-}$). These two horizons degenerate for $q=q_{\mathrm{ext}}$.
The behaviors of the metric function $f(r)$ versus $r$ for different
topologies of horizon are depicted in Fig. \ref{fig1}.

Up to now, we have obtained the higher-dimensional black hole solutions in
the context of massive gravity and in the presence of power-law Maxwell
gauge field. In the next section, we will study the thermodynamics of the
obtained solutions. To do that, we shall obtain the Smarr-type formula and
check the satisfaction of the first law of black holes thermodynamics.

\section{THERMODYNAMICS OF MASSIVE GRAVITY \label{Therm}}

The main purpose of this section is to examine the first law of
thermodynamics for massive black holes with power-law Maxwell field. It was
shown that the entropy of black holes in massive gravity still obeys the
area law \cite{cai}. It is easy to show that the entropy of black hole per
unit volume $\omega _{n-1}$ as an extensive quantity of thermodynamics is
given by \cite{cai}%
\begin{equation}
S=\frac{r_{+}^{n-1}}{4},  \label{entropy}
\end{equation}%
which is a quarter of the event horizon area \cite{cai,bek}. The electric
charge of black hole per unit volume $\omega _{n-1}$ can be calculated
through the use of Gauss law%
\begin{equation}
Q=\frac{\,{1}}{4\pi }\int r^{n-1}\left( -\mathcal{F}\right) ^{p-1}F_{\mu \nu
}n^{\mu }u^{\nu }dr,  \label{chdef}
\end{equation}%
where $n^{\mu }$ and $u^{\nu }$are respectively the unit spacelike and
timelike normals to the hypersurface of radius $r$ defined by%
\begin{equation}
n^{\mu }=\frac{1}{\sqrt{-g_{tt}}}dt=\frac{1}{\sqrt{f(r)}}dt,\text{ \ \ \ \ }%
u^{\nu }=\frac{1}{\sqrt{g_{rr}}}dr=\sqrt{f(r)}dr.
\end{equation}%
Thus, one can obtain%
\begin{equation}
Q=\frac{2^{p-1}q^{2p-1}}{4\pi }.  \label{charge}
\end{equation}%
In order to obtain the mass of black holes in massive gravity one can apply
the Hamiltonian approach presented in Ref. \cite{cai}. The total mass ($M$)
of massive black hole per unit volume $\omega _{n-1}$ can be calculated as 
\cite{cai} 
\begin{equation}
M=\frac{(n-1)m_{0}}{16\pi },  \label{Mass}
\end{equation}%
where $m_{0}$ as a function of the horizon radius $r_{+}$ was given in Eq. (%
\ref{ZM}). In order to check the first law of thermodynamic, we need to
compute Smarr-type formula for mass $M$ as a function of extensive
quantities entropy and electric charge. Using relations (\ref{entropy}), (%
\ref{charge}) and (\ref{Mass}), one can obtain the Smarr-type formula for
mass as%
\begin{eqnarray}
M(S,Q) &=&\frac{k(n-1)(4S)^{(n-2)/(n-1)}}{16\pi }-\frac{\Lambda \left(
4S\right) ^{n/(n-1)}}{8\pi n}+\frac{Q^{2p/(2p-1)}(2p-1)^{2}}{2(n-2p)\left(
4S\right) ^{\frac{n-2p}{(n-1)(2p-1)}}}\left( \frac{\pi }{2^{p-3}}\right)
^{1/(2p-1)}  \notag \\
&&+\frac{c_{0}m^{2}S}{4\pi }\left( c_{1}+\frac{(n-1)c_{0}c_{2}}{\left(
4S\right) ^{1/(n-1)}}+\frac{(n-1)(n-2)c_{0}^{2}c_{3}}{\left( 4S\right)
^{2/(n-1)}}+\frac{(n-1)(n-2)(n-3)c_{0}^{3}c_{4}}{\left( 4S\right) ^{3/(n-1)}}%
\right) .  \label{Smarr}
\end{eqnarray}%
Now, one can show that the thermodynamic quantities satisfy the first law of
thermodynamic as%
\begin{equation}
dM=TdS+UdQ,  \label{TFL}
\end{equation}%
in which%
\begin{equation}
T=\left( \frac{\partial M}{\partial S}\right) _{Q}\text{ \ \ \ \ and \ \ \ \ 
}U=\left( \frac{\partial M}{\partial Q}\right) _{S},  \label{intqua}
\end{equation}%
provided $C=p$ in (\ref{Poten}). As it is clear, for linear Maxwell case ($%
p=1$), the constant $C$\ is reduced to $1$. In the remainder of this work,
we study the effect of power-law Maxwell electrodynamics on the holographic
conductivity of dual systems with and without translational symmetry. 
\begin{figure*}[t]
\begin{center}
\begin{minipage}[b]{0.32\textwidth}\begin{center}
       \subfigure[~$n=3$]{
                \label{fig2a}\includegraphics[width=\textwidth]{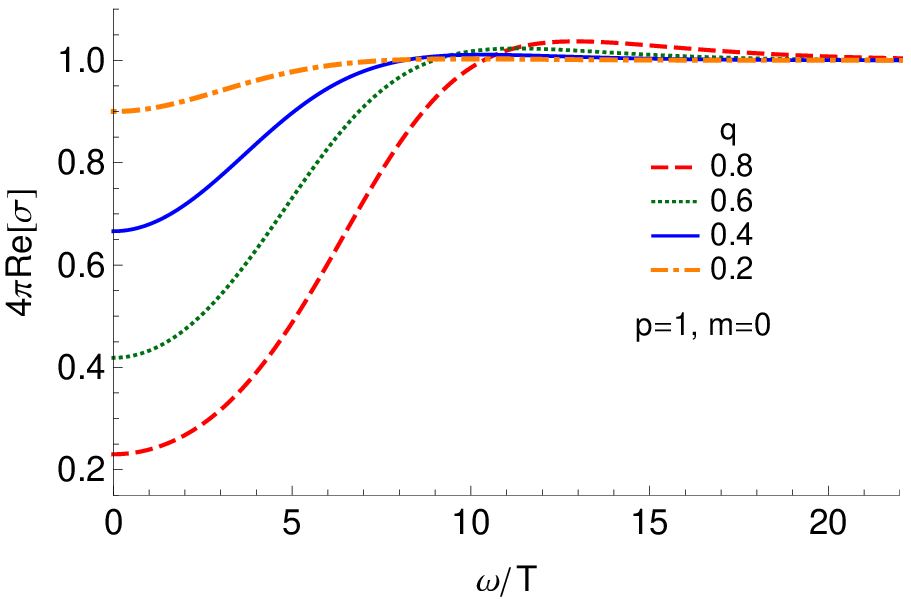}\qquad}
    \end{center}\end{minipage}\hskip+0cm 
\begin{minipage}[b]{0.32\textwidth}\begin{center}
        \subfigure[~$n=4$]{
                 \label{fig2b}\includegraphics[width=\textwidth]{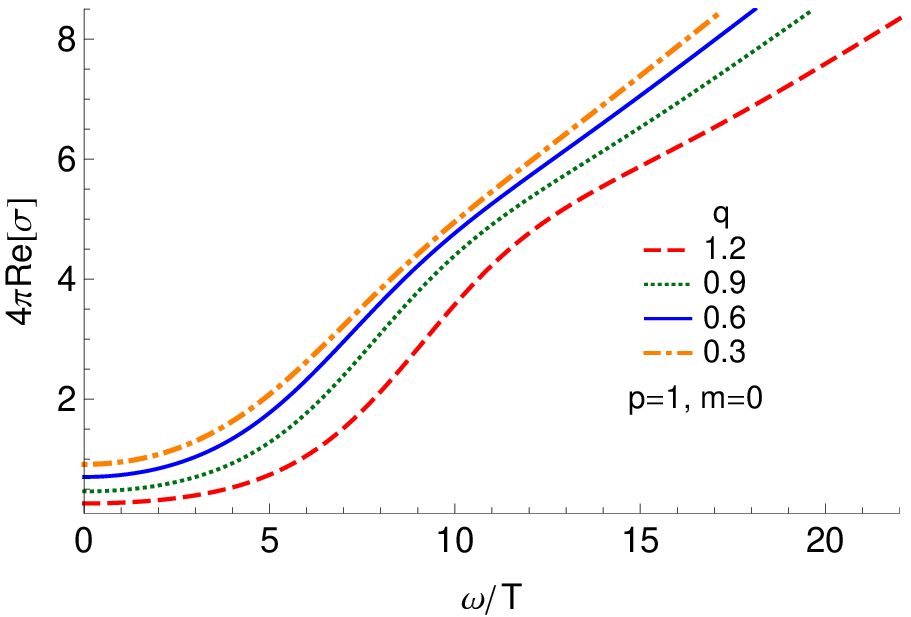}\qquad}
    \end{center}\end{minipage}\hskip0cm 
\begin{minipage}[b]{0.32\textwidth}\begin{center}
         \subfigure[~$n=3$]{
                  \label{fig2c}\includegraphics[width=\textwidth]{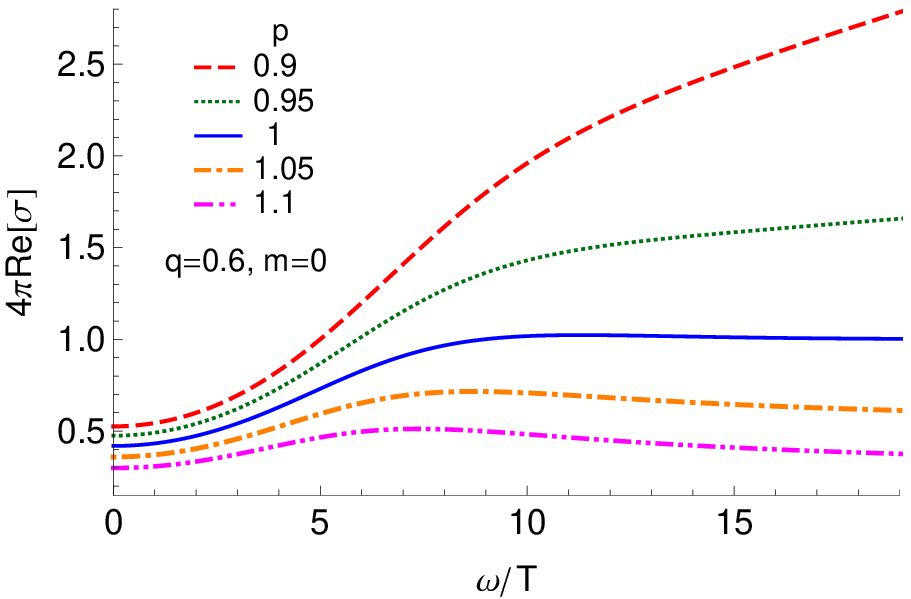}\qquad}
    \end{center}\end{minipage}\hskip0cm 
\begin{minipage}[b]{0.32\textwidth}\begin{center}
        \subfigure[~$n=4$]{
                 \label{fig2d}\includegraphics[width=\textwidth]{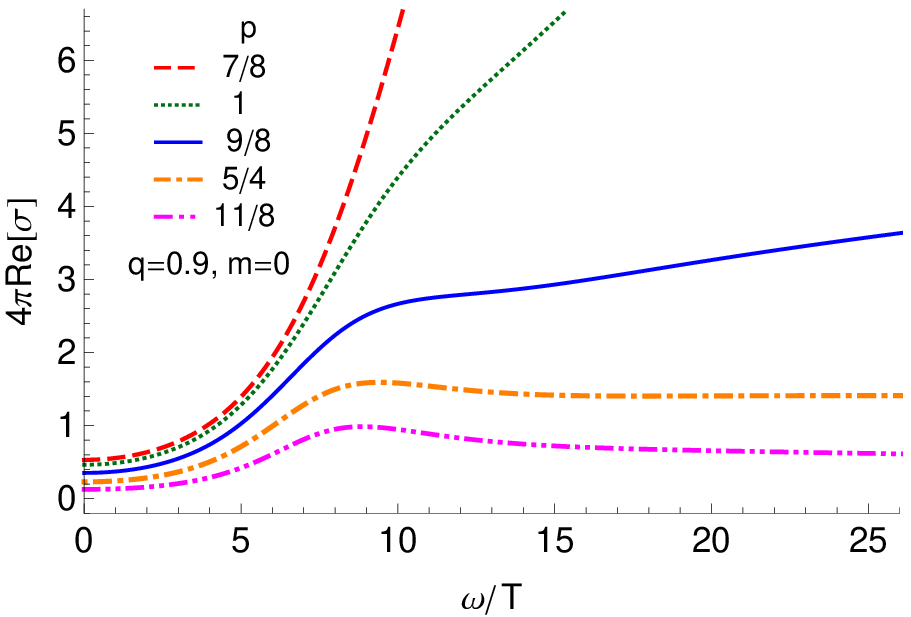}\qquad}
    \end{center}\end{minipage}\hskip+0cm 
\begin{minipage}[b]{0.32\textwidth}\begin{center}
       \subfigure[~$p=(n+1)/4$]{
          \label{fig2e}\includegraphics[width=\textwidth]{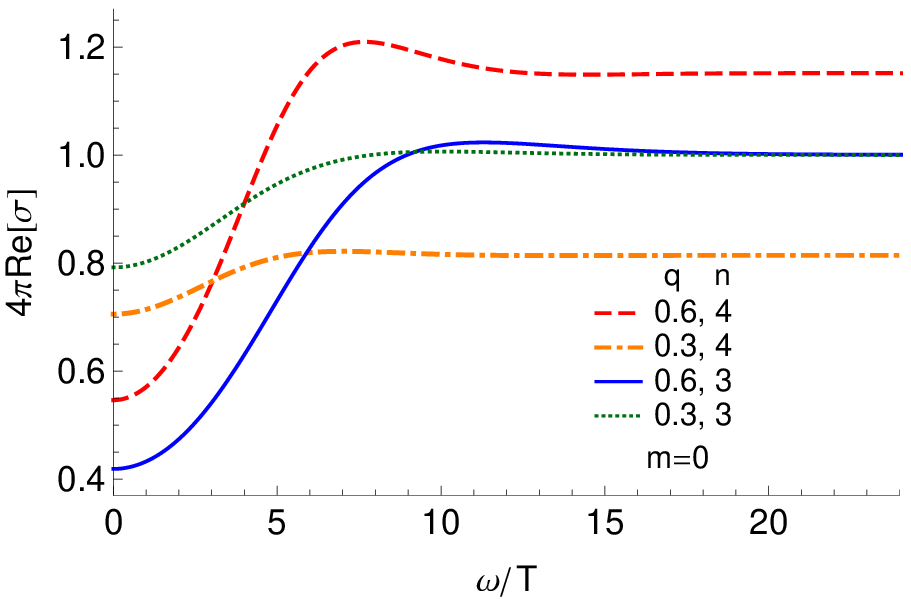}\qquad}
    \end{center}\end{minipage}\hskip0cm
\end{center}
\caption{The behaviors of real parts of conductivity $\protect\sigma $
versus $\protect\omega /T$ for $m=0$ with $l=r_{+}=1$.}
\label{fig2}
\end{figure*}

\begin{figure*}[t]
\begin{center}
\begin{minipage}[b]{0.32\textwidth}\begin{center}
       \subfigure[~$n=3$]{
                \label{fig3a}\includegraphics[width=\textwidth]{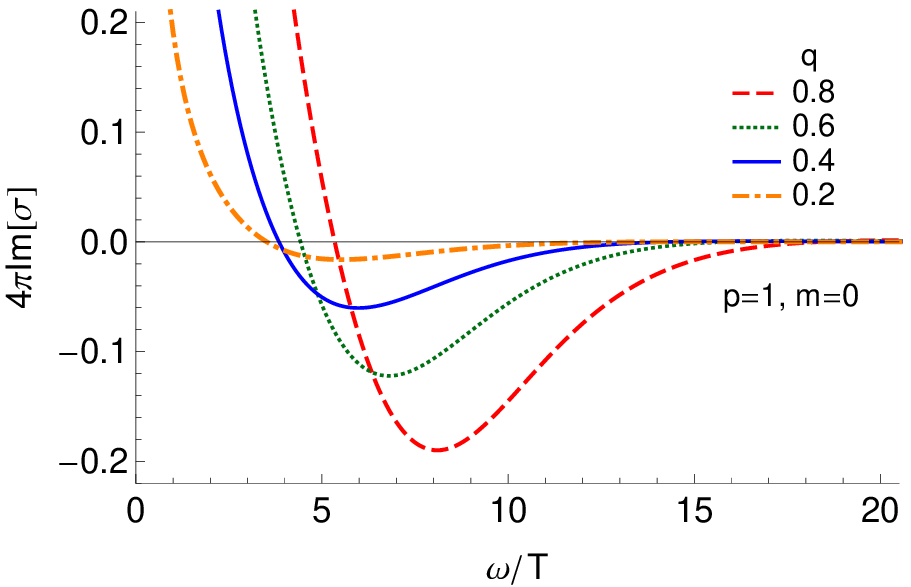}\qquad}
    \end{center}\end{minipage}\hskip+0cm 
\begin{minipage}[b]{0.32\textwidth}\begin{center}
        \subfigure[~$n=4$]{
                 \label{fig3b}\includegraphics[width=\textwidth]{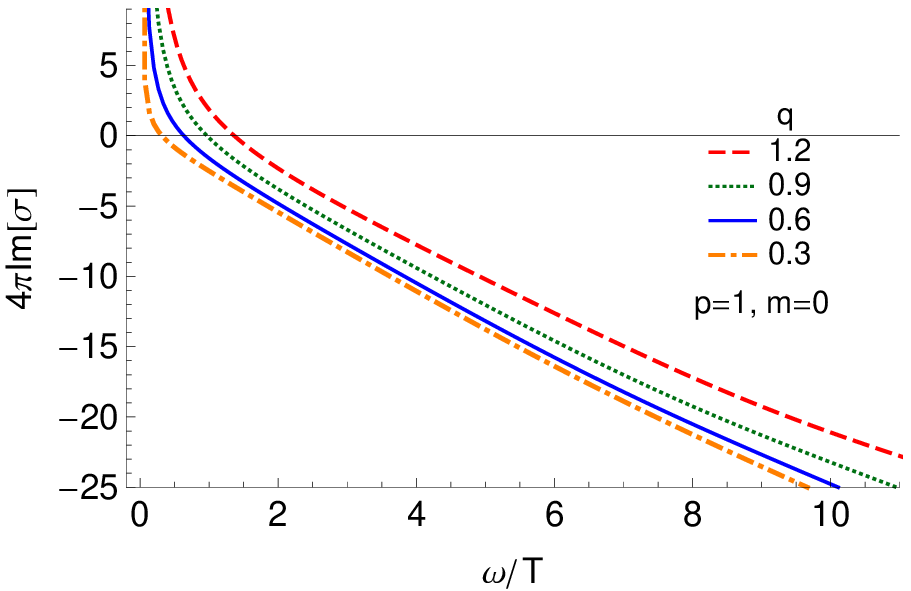}\qquad}
    \end{center}\end{minipage}\hskip0cm 
\begin{minipage}[b]{0.32\textwidth}\begin{center}
         \subfigure[~$n=3$]{
                  \label{fig3c}\includegraphics[width=\textwidth]{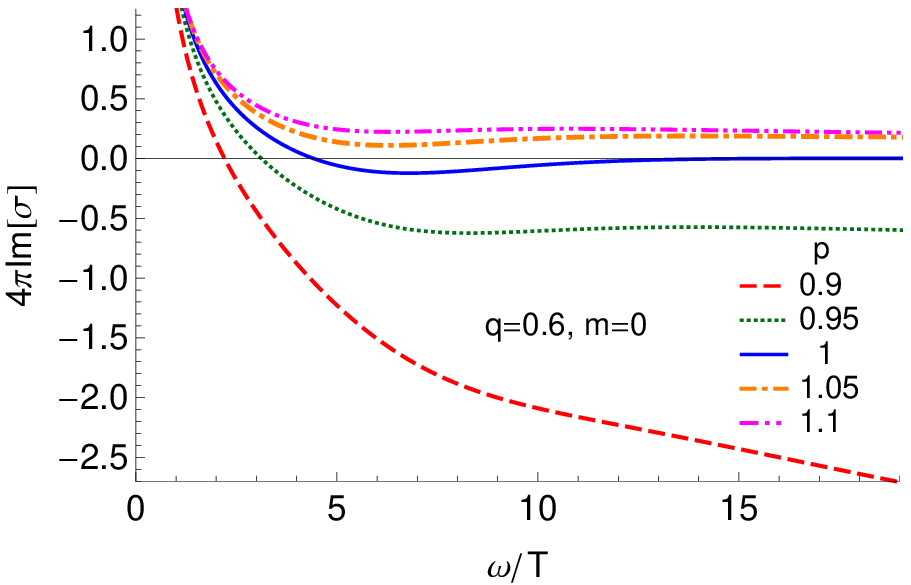}\qquad}
    \end{center}\end{minipage}\hskip0cm 
\begin{minipage}[b]{0.32\textwidth}\begin{center}
        \subfigure[~$n=4$]{
                 \label{fig3d}\includegraphics[width=\textwidth]{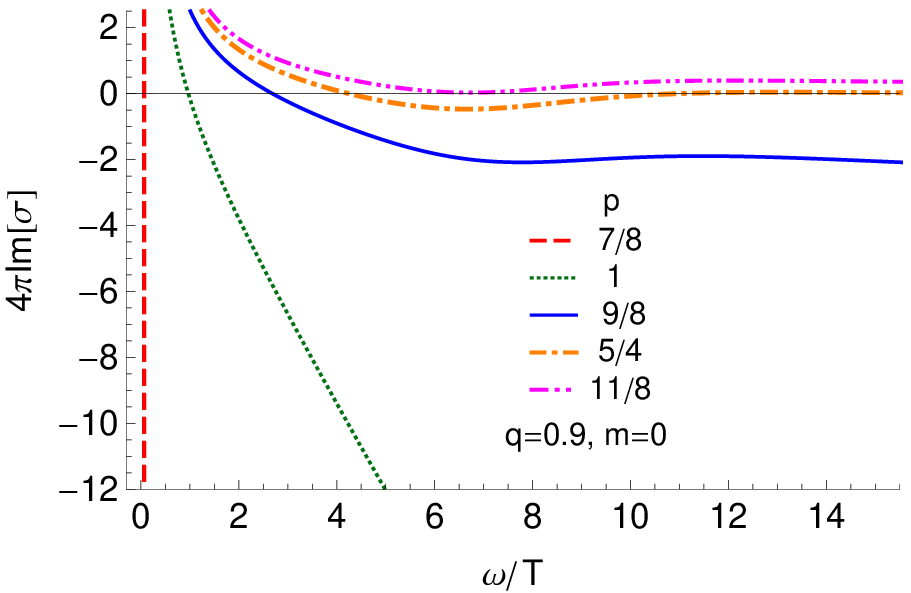}\qquad}
    \end{center}\end{minipage}\hskip+0cm 
\begin{minipage}[b]{0.32\textwidth}\begin{center}
       \subfigure[~$p=(n+1)/4$]{
          \label{fig3e}\includegraphics[width=\textwidth]{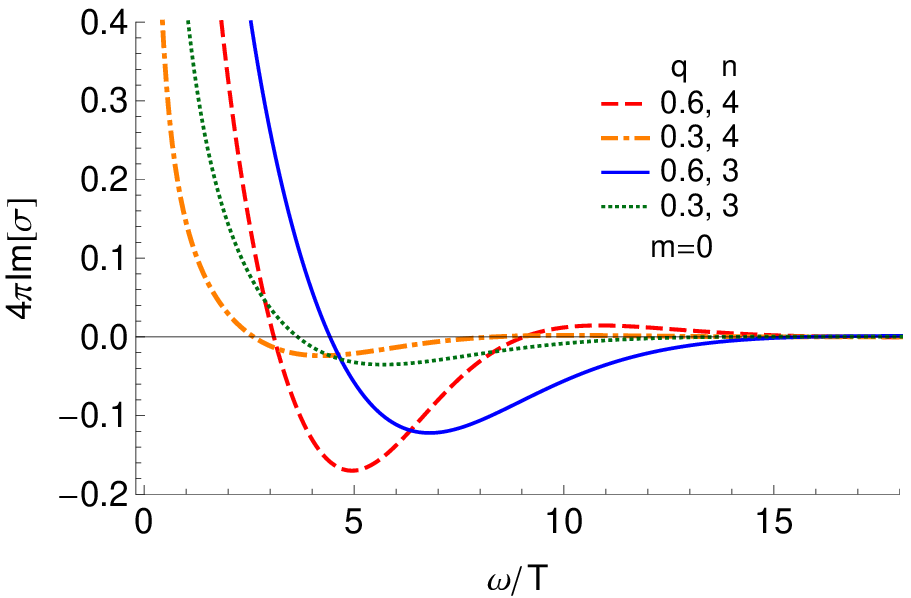}\qquad}
    \end{center}\end{minipage}\hskip0cm
\end{center}
\caption{The behaviors of imaginary parts of conductivity $\protect\sigma $
versus $\protect\omega /T$ for $m=0$ with $l=r_{+}=1$. }
\label{fig3}
\end{figure*}

\begin{figure*}[t]
\begin{center}
\begin{minipage}[b]{0.32\textwidth}\begin{center}
       \subfigure[~$n=3$]{
                \label{fig4a}\includegraphics[width=\textwidth]{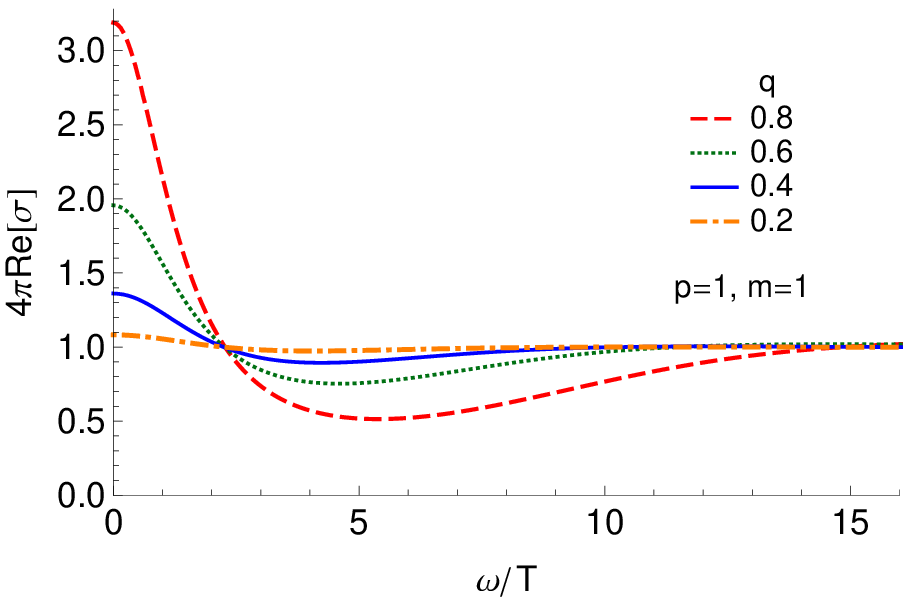}\qquad}
    \end{center}\end{minipage}\hskip+0cm 
\begin{minipage}[b]{0.32\textwidth}\begin{center}
        \subfigure[~$n=4$]{
                 \label{fig4b}\includegraphics[width=\textwidth]{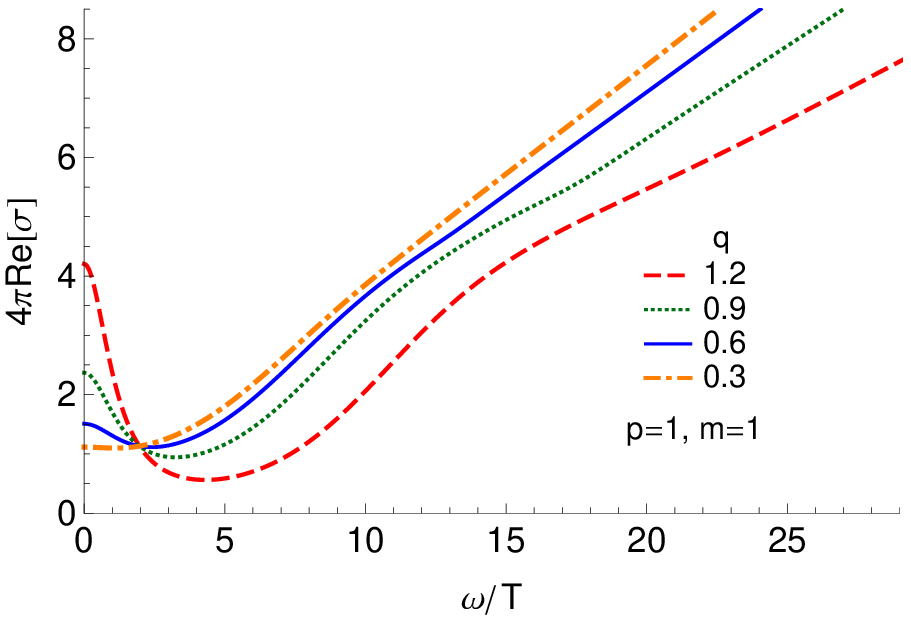}\qquad}
    \end{center}\end{minipage}\hskip0cm 
\begin{minipage}[b]{0.32\textwidth}\begin{center}
         \subfigure[~$n=3$]{
                  \label{fig4c}\includegraphics[width=\textwidth]{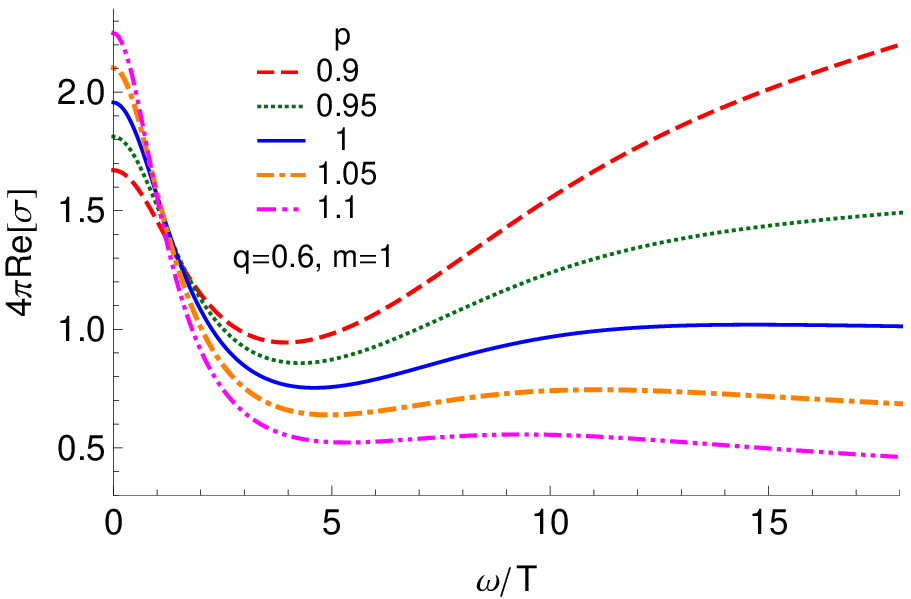}\qquad}
    \end{center}\end{minipage}\hskip0cm 
\begin{minipage}[b]{0.32\textwidth}\begin{center}
        \subfigure[~$n=4$]{
                 \label{fig4d}\includegraphics[width=\textwidth]{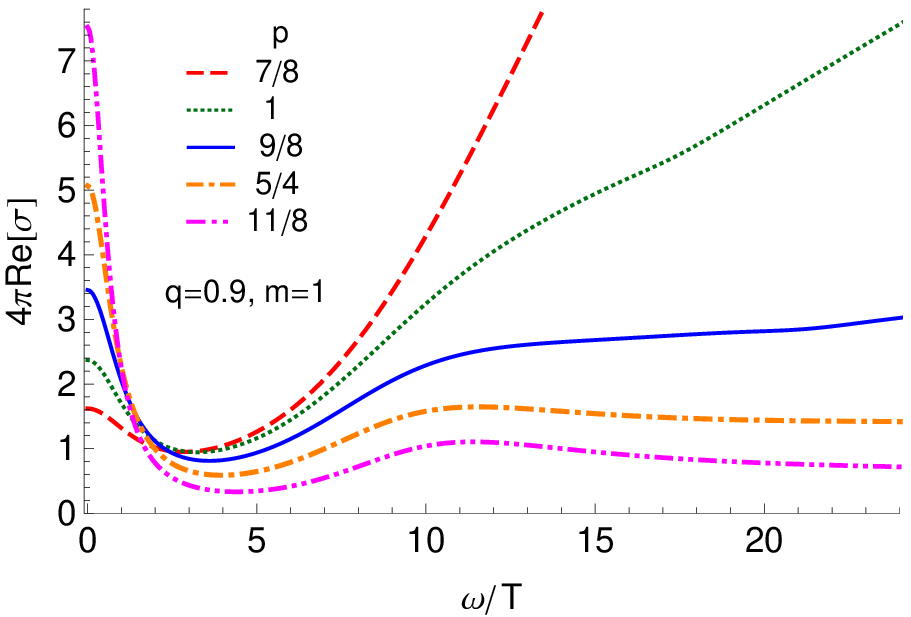}\qquad}
    \end{center}\end{minipage}\hskip+0cm 
\begin{minipage}[b]{0.32\textwidth}\begin{center}
       \subfigure[~$p=(n+1)/4$]{
          \label{fig4e}\includegraphics[width=\textwidth]{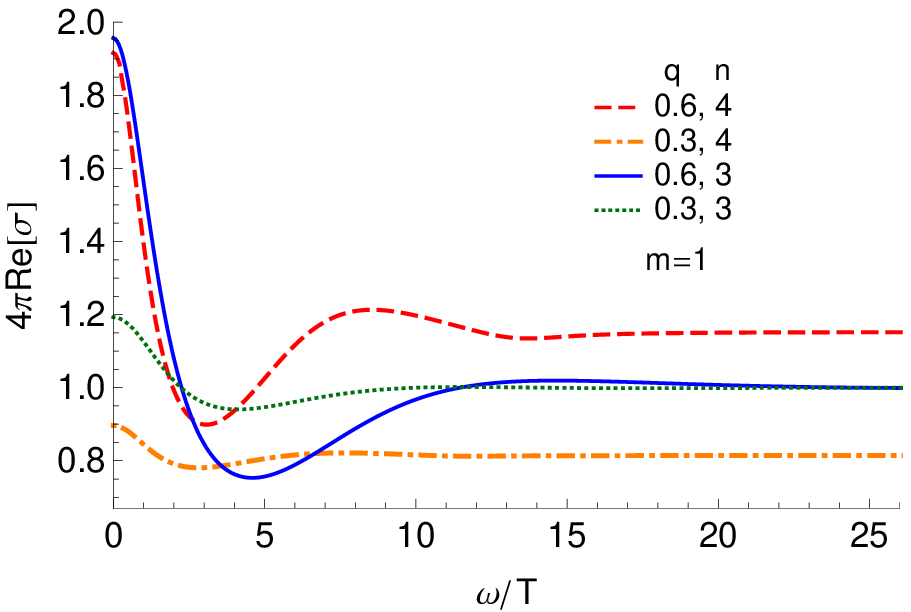}\qquad}
    \end{center}\end{minipage}\hskip0cm
\end{center}
\caption{The behaviors of real parts of conductivity $\protect\sigma $
versus $\protect\omega /T$ for $m=1$ with $l=r_{+}=1$, $c_{0}=1$, $c_{1}=-1$
and $c_{2}=0$. }
\label{fig4}
\end{figure*}

\begin{figure*}[t]
\begin{center}
\begin{minipage}[b]{0.32\textwidth}\begin{center}
       \subfigure[~$n=3$]{
                \label{fig5a}\includegraphics[width=\textwidth]{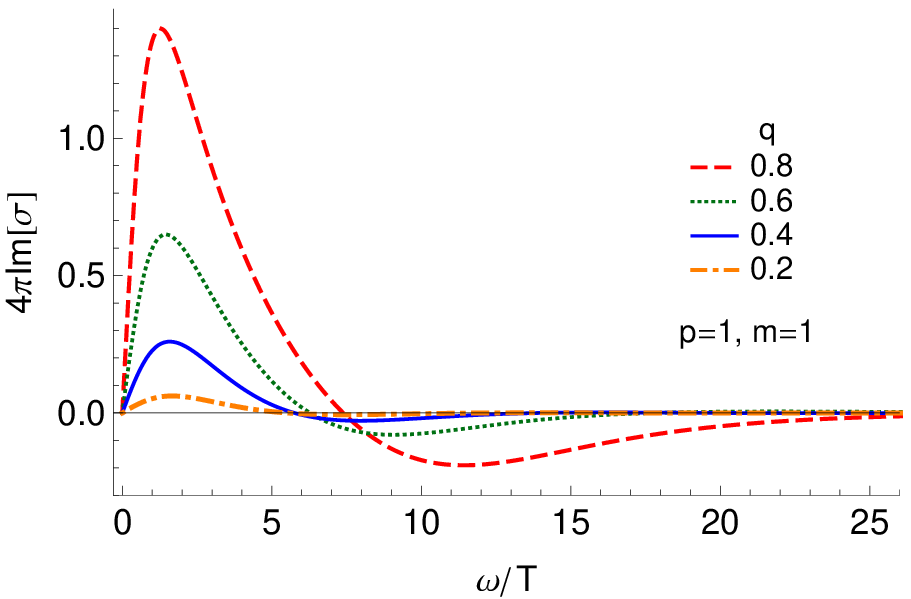}\qquad}
    \end{center}\end{minipage}\hskip+0cm 
\begin{minipage}[b]{0.32\textwidth}\begin{center}
        \subfigure[~$n=4$]{
                 \label{fig5b}\includegraphics[width=\textwidth]{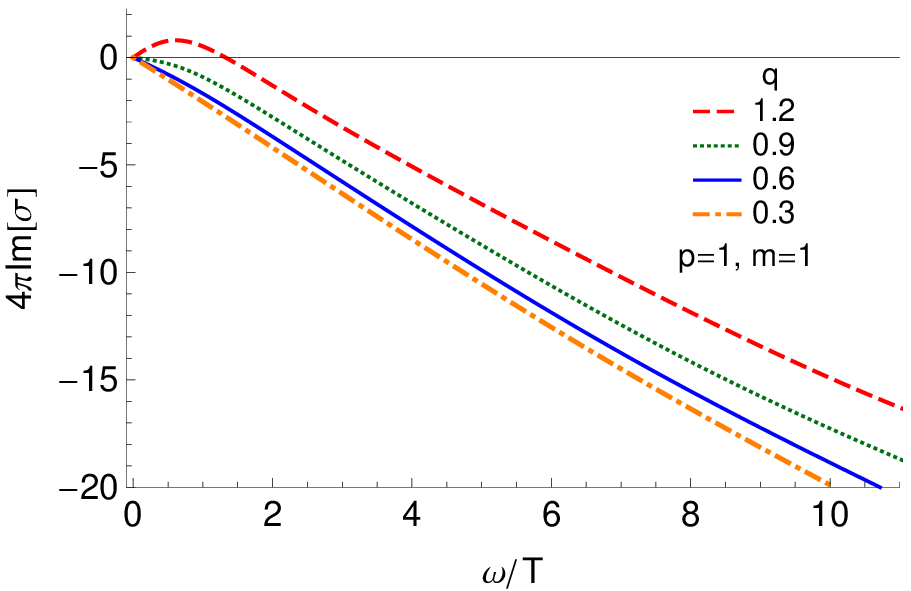}\qquad}
    \end{center}\end{minipage}\hskip0cm 
\begin{minipage}[b]{0.32\textwidth}\begin{center}
         \subfigure[~$n=3$]{
                  \label{fig5c}\includegraphics[width=\textwidth]{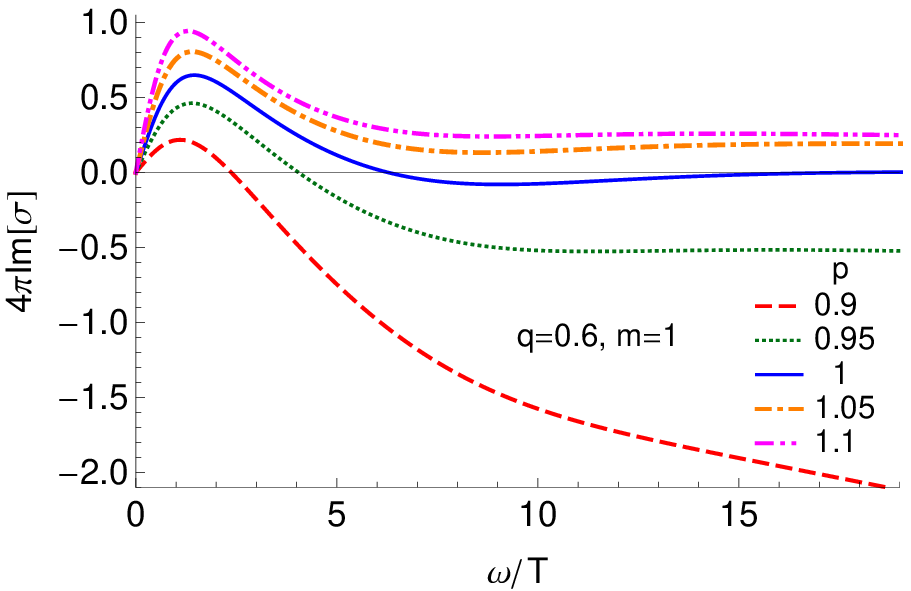}\qquad}
    \end{center}\end{minipage}\hskip0cm 
\begin{minipage}[b]{0.32\textwidth}\begin{center}
        \subfigure[~$n=4$]{
                 \label{fig5d}\includegraphics[width=\textwidth]{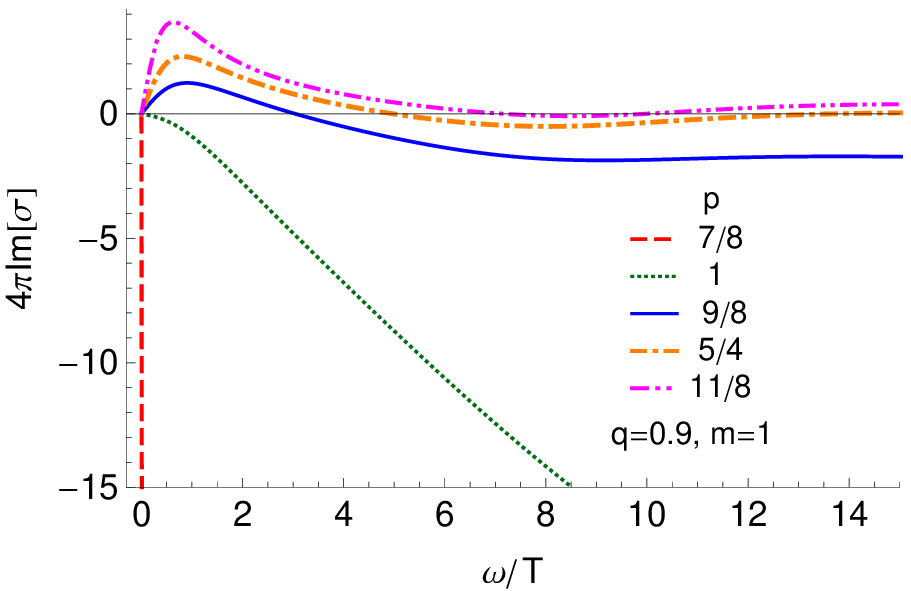}\qquad}
    \end{center}\end{minipage}\hskip+0cm 
\begin{minipage}[b]{0.32\textwidth}\begin{center}
       \subfigure[~$p=(n+1)/4$]{
          \label{fig5e}\includegraphics[width=\textwidth]{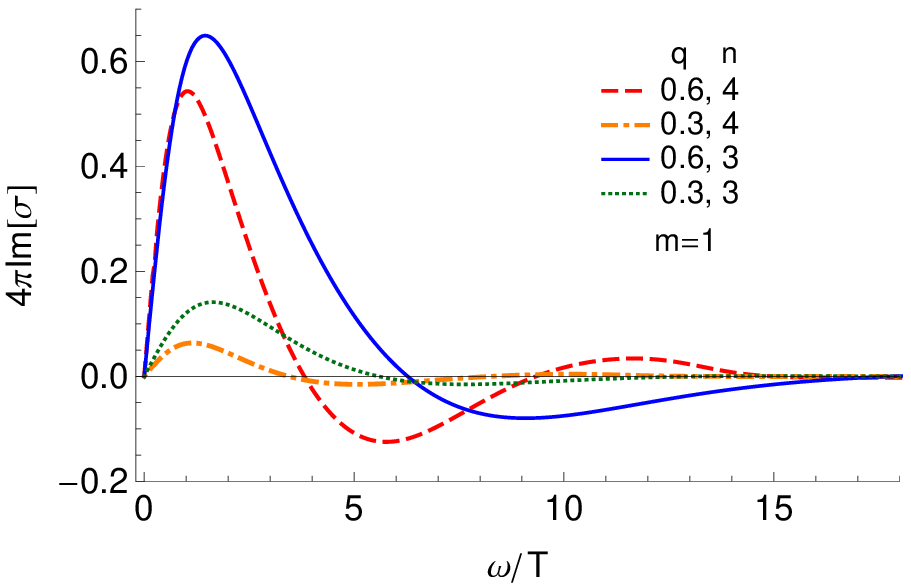}\qquad}
    \end{center}\end{minipage}\hskip0cm
\end{center}
\caption{The behaviors of imaginary parts of conductivity $\protect\sigma $
versus $\protect\omega /T$ for $m=1$ with $l=r_{+}=1$, $c_{0}=1$, $c_{1}=-1$
and $c_{2}=0$. }
\label{fig5}
\end{figure*}

\section{Holographic conductivity\label{conductivity}}

In this section, we will obtain the electrical transport behavior of the
dual field theory in the presence of a power-law Maxwell gauge field. In
order to do this, one should use the solution of the black brane ($k=0$)
found in the pervious section. First, we investigate the effects of the
power-law Maxwell electrodynamics on the holographic conductivity of dual
systems in which momentum is conserved ($m=0$). Next, we consider the
solutions dual to the systems which no longer possess momentum conservation (%
$m\neq 0$).

\subsection{Vanishing $m$}

The planer ($n+1$)-dimensional metric can be rewritten as%
\begin{equation}
ds^{2}=-\mathcal{F}(u)dt^{2}+l^{2}\mathcal{F}%
(u)^{-1}u^{-4}du^{2}+l^{2}u^{-2}\sum_{i=1}^{n-1}dx_{i}^{2},
\end{equation}%
which is given by defining $u=lr^{-1}$ in the metric (\ref{Metric}).
Accordingly, the event horizon of black brane is at $u_{+}=lr_{+}^{-1}$ and
the $n$-dimensional thermal field theory lives at $u=0$. The metric function
of spacetime in absence of massive parameter is%
\begin{equation}
\mathcal{F}%
(u)=-m_{0}l^{2-n}u^{n-2}+u^{-2}+2^{p}q^{2p}(2p-1)^{2}(n-1)^{-1}(n-2p)^{-1}%
\left[ l^{-1}u\right] ^{2(np-3p+1)/(2p-1)},
\end{equation}%
obtained by substituting $r=lu^{-1}$ and $k=0$ in Eq. (\ref{f0}). Perturbing
the vector potential component $A_{x}$ and the metric component $g_{tx}$ by
turning on $a_{x}(u)e^{-i\omega t}$ and $g_{tx}(u)e^{-i\omega t}$
respectively, we can easily derive two linear equations of motion for
electrodynamics%
\begin{equation}
a_{x}^{\prime \prime }+\left( (8p-n-3)\left( 2p-1\right) ^{-1}u^{-1}+%
\mathcal{F}^{\prime }\mathcal{F}^{-1}\right) a_{x}^{\prime }+l^{2}\omega
^{2}u^{-4}\mathcal{F}^{-2}a_{x}+h^{\prime }\mathcal{F}^{-1}\left(
g_{tx}^{\prime }+2u^{-1}g_{tx}\right) =0,  \label{ax1}
\end{equation}%
and for gravity%
\begin{equation}
g_{tx}^{\prime }+2u^{-1}g_{tx}+2^{p+1}ph^{\prime }\left(
u^{4}l^{-2}h^{\prime 2}\right) ^{p-1}a_{x}=0,  \label{ax2}
\end{equation}%
where now the prime means derivative with respect to $u$ and $h(u)$ is
electric potential in the form%
\begin{equation}
h(u)=\mu +\frac{q(2p-1)u^{(n-2p)/(2p-1)}}{(n-2p)l^{(n-2p)/(2p-1)}},
\end{equation}%
which is obtained by transforming $r\rightarrow lu^{-1}$ in Eq. (\ref%
{potential}). By eliminating $g_{tx}$ between Eqs. (\ref{ax1}) and (\ref{ax2}%
), the differential equation for $a_{x}$ is%
\begin{eqnarray}
a_{x}^{\prime \prime }+\left( (8p-n-3)\left( 2p-1\right) ^{-1}u^{-1}+%
\mathcal{F}^{\prime }\mathcal{F}^{-1}\right) a_{x}^{\prime }+a_{x}\mathcal{F}%
^{-1}\left( l^{2}\omega ^{2}u^{-4}\mathcal{F}^{-1}-2^{p+1}ph^{\prime
2}\left( u^{4}l^{-2}h^{\prime 2}\right) ^{p-1}\right) &=&0.  \notag \\
&&
\end{eqnarray}%
The behavior of above relation near the boundary ($u\rightarrow 0$) is%
\begin{equation}
a_{x}^{\prime \prime }+(4p-n-1)\left( 2p-1\right) ^{-1}u^{-1}a_{x}^{\prime
}+\cdots =0,  \label{deqax}
\end{equation}%
which has the following solution%
\begin{equation}
a_{x}(u)=a_{1}+a_{2}u^{(n-2p)/(2p-1)}+\cdots ,
\end{equation}%
where $a_{1}$ and $a_{2}$ are two constant parameters. To calculate the
expectation value of current for boundary theory, we can use the following
formula \cite{hartnoll,tong}%
\begin{equation}
\left\langle J_{x}\right\rangle =\left. \frac{\partial \mathcal{L}}{\partial
\left( \partial _{u}\delta a_{x}\right) }\right\vert _{u=0},
\end{equation}%
where $\delta a_{x}=a_{x}(u)e^{-i\omega t}$ and $\mathcal{L}$ was given in
Eq. (\ref{DL}). So, it is obvious that the holographic conductivity can be
obtained as%
\begin{equation}
\sigma =\frac{\left\langle J_{x}\right\rangle }{E_{x}}=-\frac{\left\langle
J_{x}\right\rangle }{\partial _{t}\delta a_{x}}=-\frac{i\left\langle
J_{x}\right\rangle }{\omega \delta a_{x}}=\frac{2^{p-3}p(n-2p)q^{2(p-1)}a_{2}%
}{(2p-1)\pi i\omega a_{1}}.  \label{conduc}
\end{equation}

It is easy to show that the holographic conductivity (\ref{conduc}) reduces
to $\sigma =a_{2}/\left( 4\pi i\omega a_{1}\right) $ for $n=3$ and $p=1$ 
\cite{vegh,hartnoll}. In Figs. \ref{fig2a} and \ref{fig3a}, the behaviors of
real and imaginary parts of holographic conductivity for linear Maxwell case
($p=1$) are illustrated as a function of $\omega /T$\ and for various values
of the charges of black brane $q$ for $n=3$. This figure shows that the real
part of conductivity $\mathrm{Re}[\sigma ]$\ decreases as $q$\ increases
(temperature decreases) for $\omega \rightarrow 0$ (Fig. \ref{fig2a}). Our
numerical computations show that $\mathrm{Re}[\sigma ]$\ diverges at $\omega
=0$\ independent of the value of the charge parameter $q$. Also, the maximum
value of $\mathrm{Re}[\sigma ]$\ is greater for greater $q$'s. We observe
that $\mathrm{Re}[\sigma ]$\ tends to a constant for high frequencies
independent of the value of the charge parameter. Next, we turn to study
imaginary part of the conductivity $\mathrm{Im}[\sigma ]$ plotted in Fig. %
\ref{fig3a}. Imaginary part of conductivity includes a minimum for different
charges. This minimum is deeper for larger charges (lower temperatures). At $%
\omega =0$, imaginary part of conductivity $\mathrm{Im}[\sigma ]$ diverges
(Fig. \ref{fig3a}). This fact supports our numerical computation which shows
that real part of conductivity blows up at zero frequency, according to
Kramers-Kronig relation. For high frequencies, the imaginary part of
conductivity vanishes independent of the value of charge. In Figs. \ref%
{fig2b} and \ref{fig3b}, the behaviors of real and imaginary parts of
holographic conductivity for linear Maxwell in terms of frequency for
different values of black brane's charge $q$ for $n=4$ are depicted. For low
frequencies the behavior of holographic conductivity is the same as the case 
$n=3$. However, for high frequencies the behaviors are different. In $n=3$
case, the real (imaginary) part of conductivity tends to a constant for high
frequencies whereas for $n=4$ case it increases (decreases) as $\omega $
increases.

Now, we intend to study the effect of nonlinearity of the electrodynamics
(power parameter $p$ of the power-law Maxwell field) on holographic
conductivity. Figs. \ref{fig2c}, \ref{fig2d}, \ref{fig3c} and \ref{fig3d}
show the behavior of $\mathrm{Re}[\sigma ]$\ and $\mathrm{Im}[\sigma ]$\ as
a function of $\omega /T$\ for different values of $p$\ (restricted by $%
1/2<p<n/2$) for $n=3$ and $4$. In the $\omega \rightarrow 0$ limit,
increasing $p$ leads to the smaller $\mathrm{Re}[\sigma ]$. For high
frequencies, $\mathrm{Re}[\sigma ]$\ increases (decreases) as linear
function of $\omega /T$\ and its slope increases (decreases) as $p$
decreases (increases) for $p<(n+1)/4$\ ($p>(n+1)/4$). For $p=(n+1)/4$, $%
\mathrm{Re}[\sigma ]$ and $\mathrm{Im}[\sigma ]$ tend to a constant for high
frequencies as one can see in Figs. \ref{fig2e} and \ref{fig3e}. Above
behaviors show that for high frequencies \textrm{Re}$\left[ \sigma \right]
\propto \omega ^{a}$ where $a\propto n+1-4p$. This result is important from
holographic point of view since similar results can be found in experimental
observations \cite{7,8}. In \cite{7}, for a ($2+1$)-dimensional graphene
system, it was reported that the value of \textrm{Re}$[\sigma ]$\ tends to a
constant for large frequencies. We observed such a behavior in the
conformally invariant case, $p=(n+1)/4$. For conductivity of a ($2+1$%
)-dimensional single-layer graphene induced by mild oxygen plasma exposure,
a positive slope with respect to frequency for high frequencies has been
reported in \cite{8}. We observed similar behavior for conductivity in case
of $p<(n+1)/4$. For all values of $p$, we see that $\mathrm{Im}[\sigma ]$
blows up at zero frequency (Figs. \ref{fig3c} and \ref{fig3d}). For high
frequencies, imaginary part of conductivity decreases for low values of $p$,
whereas it flattens for bigger $p$'s.

\subsection{Nonvanishing $m$}

Now, we intend to demonstrate the influence of power-law Maxwell parameter $%
p $ on the holographic conductivity in massive gravity theory. Employing
again $r\rightarrow lu^{-1}$ and setting $k=0$ in (\ref{Metricfunction}), we
obtain%
\begin{eqnarray}
\mathcal{F}(u)
&=&-m_{0}l^{2-n}u^{n-2}+u^{-2}+2^{p}q^{2p}(2p-1)^{2}(n-1)^{-1}(n-2p)^{-1}%
\left[ l^{-1}u\right] ^{2(np-3p+1)/(2p-1)}  \notag \\
&&+(n-1)^{-1}c_{0}m^{2}lu^{-1}\left(
c_{1}+(n-1)l^{-1}c_{0}c_{2}u+(n-1)(n-2)l^{-2}c_{0}^{2}c_{3}u^{2}+(n-1)(n-2)(n-3)l^{-3}c_{0}^{3}c_{4}u^{3}\right) .
\notag \\
&&
\end{eqnarray}%
Hereon,we should perturb the gauge field and the metric by turning on $%
a_{x}(u)e^{-i\omega t}$, $g_{tx}(u)e^{-i\omega t}$ and $g_{ux}(u)e^{-i\omega
t}$. At the linear regime, we have three independent differential equations
for gauge field%
\begin{equation}
\left( \mathcal{F}a_{x}^{\prime }\right) ^{\prime }+(8p-n-3)\left(
2p-1\right) ^{-1}u^{-1}\mathcal{F}a_{x}^{\prime }+l^{2}\omega ^{2}u^{-4}%
\mathcal{F}^{-1}a_{x}+h^{\prime }\left( g_{tx}^{\prime
}+2u^{-1}g_{tx}+i\omega g_{ux}\right) =0,  \label{axm}
\end{equation}%
and for massive gravity 
\begin{equation}
g_{tx}^{\prime }+2u^{-1}g_{tx}+i\omega g_{ux}+2^{p+1}ph^{\prime }\left(
u^{4}l^{-2}h^{\prime 2}\right) ^{p-1}a_{x}+ic_{0}l^{-2}\omega ^{-1}u^{2}\Xi 
\mathcal{F}g_{ux}=0,  \label{gtx}
\end{equation}%
\begin{equation}
g_{tx}^{\prime \prime }+(5-n)u^{-1}g_{tx}^{\prime
}-2(n-2)u^{-2}g_{tx}+i\omega g_{ux}^{\prime }+2^{p+1}ph^{\prime }\left(
u^{4}l^{-2}h^{\prime 2}\right) ^{p-1}a_{x}^{\prime }-i(n-3)\omega
u^{-1}g_{ux}+c_{0}\Xi u^{-2}\mathcal{F}^{-1}g_{tx}=0,  \label{gux}
\end{equation}%
in which%
\begin{equation}
\Xi =m^{2}\left(
c_{1}lu^{-1}+2(n-2)c_{0}c_{2}+3(n-3)(n-2)c_{0}^{2}c_{3}l^{-1}u+4(n-4)(n-3)(n-2)c_{0}^{3}c_{4}l^{-2}u^{2}\right) .
\end{equation}%
Eliminating $g_{tx}$ between Eqs. (\ref{axm}), (\ref{gtx}) and (\ref{gux}),
one arrives at the two following second-order differential equations%
\begin{eqnarray}
\left( \mathcal{F}a_{x}^{\prime }\right) ^{\prime }+ &&(8p-n-3)\left(
2p-1\right) ^{-1}u^{-1}\mathcal{F}a_{x}^{\prime }  \notag \\
+ &&\left[ l^{2}\omega ^{2}u^{-4}\mathcal{F}^{-1}-2^{p+1}ph^{\prime 2}\left(
u^{4}l^{-2}h^{\prime 2}\right) ^{p-1}\right] a_{x}-ic_{0}l^{-2}\omega ^{-1}%
\mathcal{F}h^{\prime }\Xi u^{2}g_{ux}=0,  \label{Eqax}
\end{eqnarray}

\begin{eqnarray}
l^{-2}u^{-2}\left( u^{4}\Xi ^{-1}\mathcal{F}\left( u^{2}\Xi \mathcal{F}%
g_{ux}\right) ^{\prime }\right) ^{\prime }-i2^{p+1}p\omega c_{0}^{-1}u^{-2}%
\left[ \Xi ^{-1}u^{4}\mathcal{F}a_{x}\left( h^{\prime }\left(
u^{4}l^{-2}h^{\prime 2}\right) ^{p-1}\right) ^{\prime }\right] ^{\prime } &&
\notag \\
+i(n-3)2^{p+1}p\omega c_{0}^{-1}u^{-2}\left[ \Xi ^{-1}u^{3}\mathcal{F}%
a_{x}h^{\prime }\left( u^{4}l^{-2}h^{\prime 2}\right) ^{p-1}\right] ^{\prime
}-(n-3)l^{-2}u^{-2}(u^{5}\mathcal{F}^{2}g_{ux})^{\prime } &&  \notag \\
+\omega ^{2}g_{ux}-i2^{p+1}p\omega h^{\prime }\left( u^{4}l^{-2}h^{\prime
2}\right) ^{p-1}a_{x}+c_{0}u^{2}l^{-2}\Xi \mathcal{F}g_{ux} &=&0.
\end{eqnarray}%
One can show that the solution of differential equation (\ref{Eqax}) near
boundary ($u\rightarrow 0$) is%
\begin{equation}
a_{x}^{\prime \prime }+(4p-n-1)\left( 2p-1\right) ^{-1}u^{-1}a_{x}^{\prime
}+\cdots =0,
\end{equation}%
which is the same as (\ref{deqax}) and also the holographic conductivity has
the same form as (\ref{conduc}). To solve above differential equations
numerically, we impose incoming boundary conditions at the horizon%
\begin{equation}
a_{x}(u),\text{ }g_{ux}(u)\propto (u_{+}-u)^{-i\omega /4\pi T},
\end{equation}%
where $T$ is the Hawking temperature.

In Figs. \ref{fig4} and \ref{fig5}, we depict the holographic conductivity
for ($2+1$)- and ($3+1$)-dimensional dual systems including momentum
dissipation in the presence of linear Maxwell and nonlinear electrodynamics.
Fig. \ref{fig5} shows that the imaginary part of conductivity near zero
frequency does not have diverging behavior in the presence of momentum
dissipation. Consequently, according to Kramers-Kronig relation, the real
part of conductivity does not diverge at $\omega =0$ and includes a Drude
peak (in contrast with the case of previous subsection with no momentum
dissipation where imaginary part of conductivity blows up at zero frequency
and accordingly real part diverges there). Also, real part of DC
conductivity becomes larger as $q$\ ($p$) increases. For high frequencies,
the behaviors of real and imaginary parts of conductivity for $n=3$ and $4$
in terms of black brane charge $q$ and nonlinear parameterar $p$ are similar
to the case of previous subsection with no momentum dissipation.

\section{Closing remarks\label{Clos}}

A gravity theory called massive gravity \cite{vegh} was proposed in order to
describe a class of strongly interacting quantum field theories with broken
translational symmetry via a holographic principle. In this letter, we
consider the massive gravity theory when the gauge field is in the form of
the power-Maxwell electrodynamics. First, we derive a class of higher
dimensional topological black hole solutions of this theory. Then, we
calculate the conserved and thermodynamic quantities of the system and check
that these quantities satisfy the first law of black holes thermodynamics on
the horizon.

The main purpose of this letter is to investigate the electrical transport
behavior of the dual field theory in the presence of a power-law Maxwell
gauge field for the obtained solutions. In order to clarify the effects of
the massive gravity on the holographic conductivity, we have first
considered the holographic conductivity of the dual systems in which
momentum is conserved ($m=0$). Then, we have extended our study to the case
where translational symmetry is broken and consequently the system no longer
possess momentum conservation ($m\neq 0$). For both cases, we have plotted
the behaviour of the real and imaginary parts of the holographic
conductivity in terms of the frequency per temperature ($\omega /T$)\ for ($%
2+1$)- and ($3+1$)-dimensional dual systems. In the former case ($m=0$), we
observed that the real part of conductivity $\mathrm{Re}[\sigma ]$\ for $n=3$
decreases as $q$\ increases (temperature decreases) for $\omega \rightarrow
0 $. Besides, $\mathrm{Re}[\sigma ]$\ has a maximum which is greater for
greater charges. Also, $\mathrm{Re}[\sigma ]$ tends to a constant for high
frequencies independent of the value of charge. In addition, the imaginary
part of conductivity $\mathrm{Im}[\sigma ]$ diverges as $\omega \rightarrow
0 $. For high frequencies, the imaginary part of conductivity vanishes
independent of the value of charge. The low frequencies behavior of
holographic conductivity for $n=4$\ is the same as the case of $n=3$. For
high frequencies, in contrast with $n=3$, the real (imaginary) part of
conductivity increases (decreases) as $\mathrm{Re}[\sigma ]$ increases for $%
n=4$. Next, we explored the effect of the power-law Maxwell field on
holographic electrical transport. We observed that increasing $p$\ leads to
the smaller $\mathrm{Re}[\sigma ]$ for $\omega \rightarrow 0$\ while for
high frequencies $\mathrm{Re}\left[ \sigma \right] \propto \omega ^{a}$\
where $a\propto (n+1-4p)$. Similar results for high frequencies can be found
in experimental observations on ($2+1$)-dimensional graphene systems \cite%
{7,8}. This is important from holographic point of view.

In the latter case ($m\neq 0$), we find out that the imaginary part of the
DC conductivity, $\mathrm{Im}[\sigma ]$, is zero at $\omega =0$ and\ becomes
larger as $q$\ increases (temperature decreases). This is in contrast to the
case without momentum dissipation. It also has a maximum value for $\omega
\neq 0$ which increases with increasing $q$ (with fixed $p$) or increasing $%
p $ (with fixed $q$) for $n=3$. For the real part of the conductivity, $%
\mathrm{Re}[\sigma ]$, we see that in case of $p=1$ the maximum value (Drude
peak) achieves at $\omega =0$. Again this is in contrast to the former case (%
$m=0$) in which the minimum value of $\mathrm{Re}[\sigma ]$, occurs for $%
\omega \rightarrow 0$. For different values of the power parameter, $p$, the
real and imaginary part of the conductivity has relative minimum and
maximum, respectively. Finally, we observed that both real and imaginary
parts of the holographic conductivity are similar to the previous case for
high frequencies.

In this work, we obtained the conductivity by applying the linear response
theory where the electric field is treated as a probe. This may restrict the
study from fully explaining the effects of nonlinearity of electrodynamics
model. Therefore, it is an interesting issue for future researches to
consider the case where the properties of the system are functions of
electric field. In such case, nonlinear response happens. Some examples of
such studies can be found in literature in Refs. \cite%
{nonlinres0,nonlinres3,nonlinres4,nonlinres5,nonlinres1,nonlinres2}.

\begin{acknowledgments}
AD and AS thank the Research Council of Shiraz University. MKZ would like to
thank Shanghai Jiao Tong University for the warm hospitality during his
visit. This work has been financially supported by the Research Institute
for Astronomy \& Astrophysics of Maragha (RIAAM), Iran.
\end{acknowledgments}

\end{document}